\documentclass[%
 aip, jcp,
 amsmath,amssymb,
 reprint,%
]{revtex4-2}

\usepackage{color}
\usepackage{xcolor}

\usepackage{hyperref}
\usepackage{CJKutf8}
\usepackage{physics}
\usepackage{graphicx}
\usepackage{dcolumn}
\usepackage{bm}

\usepackage[utf8]{inputenc}
\usepackage[T1]{fontenc}
\usepackage{mathptmx}
\usepackage{etoolbox}

\makeatletter
\def\@email#1#2{%
 \endgroup
 \patchcmd{\titleblock@produce}
  {\frontmatter@RRAPformat}
  {\frontmatter@RRAPformat{\produce@RRAP{*#1\href{mailto:#2}{#2}}}\frontmatter@RRAPformat}
  {}{}
}%
\makeatother

\definecolor{mygreen}{rgb}{0.0,0.55,0.3}

\begin{document}
\begin{CJK*}{UTF8}{gbsn}

\title{Interfacial and density fluctuations in a lattice model of  motility-induced phase separation
}
\author{Liheng Yao (姚立衡)}
\affiliation{DAMTP, center for Mathematical Sciences, University of Cambridge, Wilberforce Road, Cambridge CB3 0WA, United Kingdom}

\author{Robert L. Jack}
\affiliation{DAMTP, center for Mathematical Sciences, University of Cambridge, Wilberforce Road, Cambridge CB3 0WA, United Kingdom}
\affiliation{Yusuf Hamied Department of Chemistry, University of Cambridge, Lensfield Road, Cambridge CB2 1EW, United Kingdom}

\date{\today}

\begin{abstract}
We analyze motility-induced phase separation and bubbly phase separation in a two-dimensional lattice model of self-propelled particles. We compare systems where the dense (liquid) phase has slab and droplet geometries. We find that interfacial fluctuations of the slab are well-described by capillary wave theory, despite the existence of bubbles in the dense phase. We attribute this to a separation of time scales between bubble expulsion and interfacial relaxation. We also characterize dependence of liquid and vapor densities on the curvature of the liquid droplet, as well as the density fluctuations inside the phases. The vapor phase behaves similarly to an equilibrium system, displaying a Laplace pressure effect that shifts its density, and Gaussian density fluctuations. The liquid phase has large non-Gaussian fluctuations, but this is not accompanied by a large density shift, contrary to the equilibrium case. Nevertheless, the shift of the vapor density can be used to infer an effective surface tension that appears to also quantify capillary wave fluctuations.
\end{abstract}

\maketitle
\end{CJK*}

\section{Introduction}
\label{sec:intro}

Active particles with persistent self-propulsion exhibit motility-induced phase separation (MIPS) \cite{tailleur2008statistical, cates2015motilityinduced} into dense and dilute phases.  This process resembles liquid-vapor phase separation, which occurs 
in equilibrium systems with attractive interparticle interactions.  However, MIPS is distinctive in that it can occur in systems where such interactions are purely repulsive \cite{thompson2011lattice, fily2012athermal, bialke2013microscopic, redner2013structure, levis2014clustering, bruna2022phase}.  The interfaces between dense and dilute phases in MIPS also retain signatures of the underlying non-equilibrium dynamics, including localized entropy production \cite{nardini2017entropy} and reverse Ostwald ripening \cite{tjhung2018cluster}.

More generally, interfaces in non-equilibrium systems appear in many contexts including cellular tissues \cite{sepulveda2013collective, brugues2014forces, alert2020physical}, bacterial colonies \cite{harshey2003bacterial, hallatschek2007genetic} and bird flocks \cite{bialek2012statistical}.  MIPS systems may be regarded as minimal models for interfaces between such active phases, and their theoretical description remains an important challenge.  Recent numerical work has also shown that models that undergo MIPS display non-trivial interfacial phenomena when interacting with fixed walls, such as capillary action \cite{wysocki2020capillary} and wetting \cite{turci2021wetting, turci2024partial} in the absence of attractive interactions.

For liquid-vapor interfaces at equilibrium, there are established theories for capillary waves and for the (Laplace) pressure difference across a curved interface (see e.g. Refs.~\onlinecite{rowlinson1982molecular, safran2003statistical}). These are both controlled by the surface tension, which also has a mechanical interpretation, via the stress tensor.  Many of the same phenomena occur in MIPS, but different aspects of the interfacial behavior are controlled by different surface-tension-like quantities.  For example, the mechanical surface tension of a MIPS interface in a system of active Brownian particles was found to be negative, despite positive measured values of the capillary wave surface tension \cite{bialke2015negative, patch2018curvaturedependent, langford2024theory}.

Significant insights on this matter have also been gained by studying phenomenological field theories that aim to capture the behavior of active systems at a hydrodynamic level \cite{tailleur2008statistical, stenhammar2013continuum, wittkowski2014scalar, speck2014effective, tjhung2018cluster, solon2018generalized, solon2018generalizeda}. In particular, in Active Model B+ (AMB+) \cite{tjhung2018cluster}, where all symmetry-allowed terms that break detailed balance up to the fourth gradient and second density orders are considered, it was observed that the effective surface tensions controlling the dynamics of Ostwald ripening \cite{tjhung2018cluster} and capillary waves \cite{fausti2021capillary} in that model can each become negative. The former negativity results in the phenomenon of reverse Ostwald ripening, where small vapor bubbles grow at the expense of larger ones, while the latter results in exotic ``active foam'' phases. 

In this work, we explore these issues in a microscopic model, which is a two-dimensional active lattice gas (ALG), inspired by Ref.~\onlinecite{mason2023exact}.  In addition to MIPS, the model also exhibits bubbly phase separation \cite{tjhung2018cluster,fausti2024statistical}, which is attributable to reverse Ostwald ripening (see also Ref.~\onlinecite{shi2020selforganized} for bubbly phases in another lattice model).  By means of extensive numerical simulations, we measure the capillary wave spectrum of the bulk liquid-vapor interface, which allows a surface tension to be extracted.  We also measure the densities of the phases in both slab and droplet geometries: these differ due to effects similar to that of Laplace pressure in equilibrium \cite{solon2018generalized}, where the density shifts are be related to the surface tension via the compressibilities of the phases, which are related in turn to bulk density fluctuations. Motivated by this observation, we also measure bulk density fluctuations in the ALG.  

To interpret these results, we discuss a simple (local) theory of the steady state probability distribution of phase-separated systems on large length scales.  This theoretical model is sufficient to describe equilibrium phase separation, but is not a priori restricted to that case.  Within this theory, we recover equilibrium-like relationships between density shifts, bulk density fluctuations and capillary waves.  For the ALG, our numerical results show that these equilibrium-like relationships are violated in general. For example,
the theory predicts that the curvature-induced density shifts of the two phases are in proportion to their bulk density fluctuations, but this is violated in numerical simulations.
This indicates that the simple local description must break down, which we attribute to long-ranged effective interactions.  Additional evidence for such interactions appears in bubbly phases, whose bubble-size distribution seems to differ between bulk states and phase-separated droplets at the same density, breaking the equilibrium equivalence of ensembles. Despite this, we find (for our parameters) that the density shift for the vapor phase numerically obeys an equilibrium-like relation involving the capillary-wave surface tension and the bulk fluctuations of the vapor. Together, these results clarify the differences between MIPS systems and liquid-vapor coexistence at equilibrium.

In the remainder of this work, Sec.~\ref{sec:algmod} introduces our ALG, Sec.~\ref{sec:algtheory} develops the theories required for our investigation, and Sec.~\ref{sec:algres} reports our results, which are then discussed in Sec.~\ref{sec:algcon}.

\section{Model definition}
\label{sec:algmod}

Our ALG is a lattice description of persistent active particles with short-ranged purely repulsive interactions, undergoing overdamped motion in contact with a heat bath of constant temperature.  It is defined on a square lattice with $(L/a)^2$ sites, where $L$ is the system size and $a$ the lattice spacing.  There are $N = \phi L^2 / a^2$ particles, and the orientation of particle $k$ is $\vb{u}_k = (\cos \theta_k, \sin \theta_k)$. The angle $\theta_k$ takes continuous values between $0$ and $2 \pi$ to avoid the lattice artifacts observed in models defined on square lattices with discrete orientations \cite{whitelam2018phase}. The state of the model evolves over time with the following rules:

\begin{enumerate}
    \item \emph{Exclusion}: jumps to occupied lattice sites are forbidden.
    \item The system has translational diffusion constant $D_E$ and persistence $v_0$: if particle $k$ is on site $i$ then it jumps to an unoccupied nearest neighbor site $j$ with rate
    \begin{equation}
        W_{ij} = \frac{2 D_E / a^2} {1 + \exp\left(- \frac{av_0}{D_E} \vb{u}_k \cdot \vb{e}_{ij}\right)}  \; ,
    \end{equation}
    where $\vb{e}_{ij}$ is the unit vector pointing from site $i$ to site $j$.
    \item \emph{Rotational diffusion}: the orientations of the particles in the system evolve with a rotational diffusion constant $D_O$.
\end{enumerate}

The particle jumps are implemented through the Gillespie algorithm \cite{gillespie1977exact, masuda2022gillespie, mason2023exact}: at each time step the total rate of all allowed transitions $R = \sum W_{ij}$ is computed, and one of the events is picked to happen with probability $W_{ij} / R$ by a binary search tree. The time is then updated by $1 / R$, and all affected rates are also updated. This process is repeated over a time increment $\delta t$, such that $a^2 D_E^{-1} \ll \delta t \ll D_O^{-1}$. After each such increment, the orientation of each particle is updated by an independent Gaussian random variable $\delta \theta_i = \mathcal{N} (0, \sqrt{2 D_O \delta t})$.

\begin{figure}
    \centering
    \includegraphics[width=1.0\linewidth]{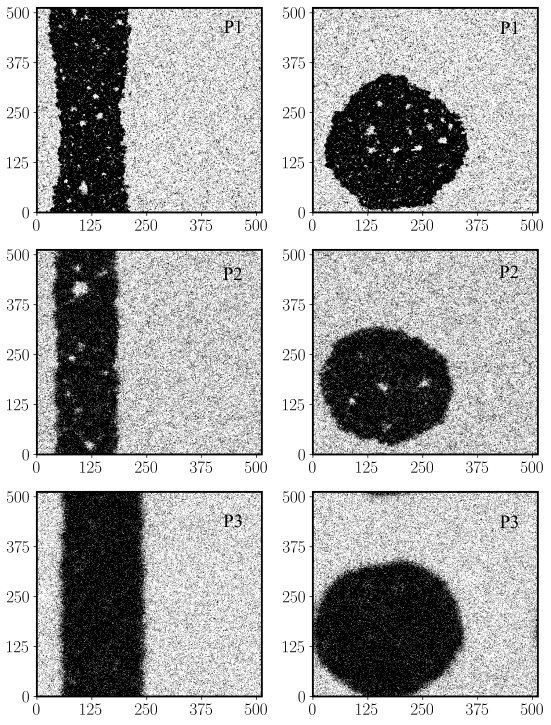}
    \caption{Snapshots of typical configurations of the ALG undergoing MIPS in the slab and droplet geometries for various parameters, at $L = 512$.}
    \label{fig:mips}
\end{figure}

The model satisfies \emph{local detailed balance} (see e.g. Ref.~\onlinecite{maes2021local}): the ratio between forward and backward transitions is 
\begin{equation}
    \frac{W_{ij}}{W_{ji}} = \exp \left(\frac{av_0}{D_E} \vb{u}_k \cdot \vb{e}_{ij} \right) \; .
    \label{eqn:ldb}
\end{equation}
The quantity in the exponent can be related to the work done by the particles' propulsive forces, which is equal to the dissipated energy in steady state.  With the aid of the Einstein relation we identify the self-propulsion force on particle $k$ as $\vb{F}_k = \vb{u}_k k_B T v_0 / D_E$, so the exponent in \eqref{eqn:ldb} is $\vb{F}_k \cdot \Delta \vb{x}_k / (k_BT)$ with $\Delta \vb{x}_k = a\vb{e}_{ij}$ the particle displacement.

In the limit $a v_0 \ll D_E$, this model reduces to the model of Ref.~\onlinecite{mason2023exact} at leading order, where the transition rate is
\begin{equation}
    W_{ij} = a^{-2} D_E + \frac{1}{2} a^{-1} v_0 \vb{u}_i \cdot \vb{e}_{ij} \; .
\end{equation}
Ref.~\onlinecite{mason2023exact} derived an exact hydrodynamic description of this model by taking $a\to0$ at fixed $L$, so the number of lattice sites goes to infinity (see also Refs.~\onlinecite{kourbane2018exact, agranov2021exact}).

We identify two important length scales that control the physical behavior of the model, which are the diffusive lengthscale $l_D = \sqrt{D_E / D_O}$ and the persistence length $l_p$, which gives the distance traveled by an isolated particle within its orientational diffusion timescale $D_O^{-1}$.  The ratio of these lengths is  the Péclet number $\mathrm{Pe} = l_p / l_D$. In terms of the bare model parameters we have
\begin{equation}
    l_p = \frac{2 D_E}{a D_O} \tanh \left( \frac{a v_0}{2 D_E} \right) \; , \quad  \mathrm{Pe} = \frac{2}{a} \sqrt{\frac{D_E}{D_O}} \tanh \left( \frac{a v_0}{2 D_E} \right) \; .
    \label{eqn:pec}
\end{equation}
For $a v_0 \ll D_E$, these parameters converge to $l_p = v_0 / D_O$, and $\mathrm{Pe} = v_0 / \sqrt{D_E D_O}$, consistent with Ref.~\onlinecite{mason2023exact}.

For the numerical implementation of the model we set $a=1$ and $D_O=1$, which set the units of length and time respectively.  We perform finite-size scaling by increasing the system size $L$ at fixed $a$, so the rate $W_{ij}$ remains constant. Note that this is different from the hydrodynamic limit~\cite{mason2023exact}, where $a\to0$ at fixed $L$.

\begin{table}
\centering
\begin{tabular}{| c | c | c | c | c | c |} 
    \hline
    \;State\;  & $v_0$ & $D_E$ & $l_p$ & $\mathrm{Pe}$ & $\phi$ \\ 
    \hline
    P1 & \;$50$ \; & \; $10$ \; & \;$19.7$\; &\; $6.24$\; & \; $0.40$ \; \\ 
    \hline
    P2 & $50$ & $50$ & $46.2$ & $6.54$ & $0.40$ \\ 
    \hline
    P3 & $110$ & $200$ & $107$ & $7.59$ & $0.45$ \\ 
    \hline
\end{tabular}
\caption{Three state points considered in this work.  [We fix $D_O=1$ and $a=1$; $l_p$ and Pe are related to $v_0,D_E$ by \eqref{eqn:pec}.]}
\label{table:params}
\end{table}

Similar to other models of persistent active matter, this ALG shows MIPS for sufficiently large persistence and density.   We choose three sets of parameters (shown in Table~\ref{table:params}), which we label as P1, P2 and P3. The system exhibits MIPS for all these parameters.
Snapshots of the system in the phase separated steady state in the slab and droplet geometries are shown in Fig.~\ref{fig:mips}.  For P1 and P2 the system shows signs of bubbly phase separation \cite{tjhung2018cluster, fausti2024statistical}, similar to the lattice model of Ref.~\onlinecite{shi2020selforganized}.  We also note that P2 is close to the boundary of stability of the MIPS state: increasing  $D_E$ from 50 to 60 reduces the Pe enough to destroy MIPS.

\section{Theory}
\label{sec:algtheory}

This Section presents the theoretical frameworks within which we analyze our numerical results.
Their applicability to equilibrium phase separation is verified in Appendix~\ref{app:ising}, which shows numerical results for an two-dimensional Ising model with conserved dynamics. Their applicability to the ALG is tested in Sec.~\ref{sec:algres}.

\subsection{Capillary wave theory}
\label{sec:cwt}

Interfacial fluctuations in equilibrium phase separation are governed by capillary wave theory (CWT) (see e.g. Ref.~\onlinecite{safran2003statistical}). {Recent work has also extended CWT to models that display MIPS \cite{patch2018curvaturedependent, fausti2021capillary, langford2024theory}.}  Here, we briefly review the main ingredients of CWT.  For the slab geometry in Fig.~\ref{fig:mips}, we consider each interface separately and we associate each point on the interface with its $y$-coordinate.  The most likely interfacial configuration is perfectly flat and we write $h_y$ for the (horizontal) deviation of the interface from the flat profile, at position $y$.
CWT applies for large-scale interfacial fluctuations and states that 
the probability (or probability density) of observing an interfacial profile $h_y$ is related to the contour length of the interface and the surface tension $\gamma_{\rm cw}$ as
\begin{equation}
        P(h_y) = P_0 \exp \left[- \gamma_{\rm cw} \int_{0}^{L} dy \left( \sqrt{1 + \left( \frac{dh_y}{dy} \right)^2} - 1 \right) \right] \\ \; ,
\end{equation}
where $P_0$ is the probability of observing a flat interface and we have absorbed a factor of $k_B T$ into $\gamma_{\rm cw}$ for compactness of notation.  For small fluctuations around a planar interface, this reduces to
\begin{equation}
      P(h_y)  \approx P_0 \exp \left[- \frac{\gamma_{\rm cw}}{2} \int_{0}^{L} dy \left( \frac{dh_y}{dy} \right)^2 \right] \; .
    \label{eqn:cwtp}
\end{equation}

To measure $\gamma_{\rm cw}$ from numerical simulations, one Fourier transforms as
\begin{equation}
    h_q = \frac{1}{L} \int_{0}^{L} dy \: h_y e^{-i q y} \; , \quad h_y = \sum_{q} h_q e^{i q y} \;  , \quad q = \frac{2 \pi n}{L} \; ,
\end{equation}
for integer $n$.  
Since $h_y$ is defined as the deviation from a flat reference profile we have $h_{q=0}=0$.
For $q\neq 0$, Eq.~\eqref{eqn:cwtp} implies
\begin{equation}
	\langle |h_q | ^2 \rangle = \frac{1}{q^2 L \gamma_{\rm cw} } \; .
	\label{eqn:cwt}
\end{equation}
We recall that the CWT is defined for large-scale fluctuations of the interface: for the lattice models considered here this means that \eqref{eqn:cwt} is relevant for $qa\ll 1$.  See Appendix~\ref{app:cwt} for a numerical example.

We also define the interfacial width $w$, which captures interfacial broadening due to capillary waves
\begin{equation}
    w^2 = \frac{1}{L} \int_{0}^{L} dy \langle h_y^2 \rangle = \sum_{q} \langle |h_q| ^2 \rangle \; .
    \label{eqn:cwtw}
\end{equation}
Using \eqref{eqn:cwt} and \eqref{eqn:cwtw} yields (for these two-dimensional systems)
\begin{equation}
    w^2 = \frac{L}{12 \gamma_{\rm cw}} \; .
    \label{eqn:wid}
\end{equation}
For large $L$, the sum over wavevectors in \ref{eqn:cwtw} is dominated by the small $q$ for which CWT holds, so measurements of $w^2$ in large systems can be used to infer $\gamma_{\rm cw}$.

\subsection{Probabilistic theory of phase separation}
\label{sec:prob}

The equilibrium theory of phase separation rests on the equality of chemical potential and pressure between the coexisting phases. In finite systems with curved interfaces, this equal-pressure criterion is modified by a Laplace pressure.  This section outlines a derivation of these conditions, based on the probability of different phase-separated configurations.  This is useful in the following since it can be applied to MIPS systems where the equilibrium concepts of pressure and chemical potential can be ambiguous \cite{solon2015pressure, speck2016ideal}.

Consider a phase separated system in a configuration with coexisting liquid and vapor of volumes $V_l,V_v$ respectively, and with $N_l,N_v$ particles in each phase.  Also $V=V_l+V_v$ and $N=N_l+N_v$.  
For large systems in two dimensions, the probability of seeing such a configuration is assumed to take the form
\begin{equation}
    P(N_l, N_v, V_l, V_v, l) \propto \exp \left[-V_l f_l \left(\frac{N_l}{V_l}\right) - V_v f_v \left(\frac{N_v}{V_v} \right) - \gamma_{\rm cw} l \right] \; .
    \label{eqn:thp}
\end{equation}
where $\gamma_{\rm cw}$ is the surface tension that appears in CWT and $l$ is the interfacial length.  The form \eqref{eqn:thp} is familiar from large deviation theory (see e.g. Ref.~\onlinecite{touchette2009large}), in which case $f_l$ is the rate function for the density of the liquid state, whose distribution becomes sharply peaked when $V_l$ is large.  In the equilibrium theory then $f_l$ would be the (Helmholtz) free energy of the liquid, per unit volume (rescaled by $k_BT$), and similarly for $f_v$.  The applicability of \eqref{eqn:thp} to non-equilibrium phase separation is not clear a priori: this will be discussed in later Sections.

For large systems, the system remains close to its most likely configuration which we can identify by maximizing the probability. First we maximize with respect to $N_l$ at fixed $N,V_l,V_v,l$ (that is, fixed geometry), which yields
\begin{equation}
    f'_l (\rho_l) = f'_v (\rho_v)  \; ,
    \label{eqn:mu}
\end{equation}
where the prime indicates differentiation with respect to the the densities of the phases, which are denoted $\rho_{l,v}=N_{l,v}/V_{l,v}$. In equilibrium this is equality of chemical potential between the phases.

Performing the same operation for $V_l$ yields an identity analogous to the equilibrium equality of pressure. In the slab configuration we keep fixed $N_l,N_v,V,l$ and obtain
\begin{equation}
     f_l (\rho_l^s) - \rho_l^s f'_l (\rho_l^s) = f_v (\rho_v^s) - \rho_v^s f'_v (\rho_v^s) \; .
    \label{eqn:ps}
\end{equation}
where $\rho_l^s$ is the liquid density in the slab geometry, and similarly $\rho_v^s$ for the vapor.

For curved interfaces, varying $V_l$ also causes a change in $l$. For a circular droplet of radius $R$ we have $dl = dV_l / R$. Maximizing the probability yields
\begin{equation}
     \frac{\gamma_{\rm cw}}{R} + f_l (\rho_l^d) - \rho_l^d f'_l (\rho_l^d) = f_v (\rho_v^d) - \rho_v^d f'_v (\rho_v^d) \; .
     \label{eqn:pd}
\end{equation}
where the densities of the phases in this droplet geometry are $\rho_{l,v}^d$. In equilibrium, $\gamma_{\rm cw}/R$ is known as the Laplace pressure.

Eq.~\eqref{eqn:mu} holds for both slab and droplet geometries, separately (with different densities in each case).  By considering the difference $ f'_l (\rho_l^d) - f'_l (\rho_l^s)$ and assuming the density difference to be small we obtain
\begin{equation}
    \alpha_l (\rho_l^d - \rho_l^s) = \alpha_v (\rho_v^d - \rho_v^s) \; .
    \label{eqn:lapcp}
\end{equation}
where 
\begin{equation}
\alpha_l = f_l''(\rho_l^s) \; , \qquad \alpha_v = f_v''(\rho_v^s) \; ,
\label{eqn:thf}
\end{equation}
are second derivatives of the free energy, which in equilibrium are inversely proportional  to the compressibilities of the phases.

Subtracting \eqref{eqn:ps} from \eqref{eqn:pd} and applying a similar argument one obtains
\begin{equation}
    \frac{\gamma_{\rm cw}}{R} = \rho_l^s \alpha_l (\rho_l^d - \rho_l^s) - \rho_v^s \alpha_v (\rho_v^d - \rho_v^s) \; .
 \end{equation}
Then using \eqref{eqn:lapcp} and rearranging yields
\begin{equation}
\rho_v^d - \rho_v^s = \frac{\gamma_{\rm cw}}{R \alpha_v (\rho_l^s - \rho_v^s)} \; .
    \label{eqn:lap}
\end{equation}
Hence we have related the density difference between slab and droplet configurations to the capillary wave surface tension.  The result is in direct analogy with the equilibrium case where this density difference is due to Laplace pressure.  Here we have derived it directly from \eqref{eqn:thp}: this does not assume that the system is in a thermal equilibrium state, but only that the log-probability of a given configuration is given by a sum of local terms, which can be separated into bulk and interfacial contributions.

In numerical simulations, the (mean) density of the phases in slab/droplet geometries can be measured by placing boxes of fixed size $V_b$ inside the bulk of the phases and counting the number of particles inside.  Since the density differences tend to be small, some care is required to avoid numerical errors (for example that the box is sufficiently far from the interfaces to measure a bulk density).  See Appendix~\ref{app:rhosd} for an example at equilibrium.  The next Section discusses independent estimates of $\alpha_{v,l}$ in which case \eqref{eqn:lap} can be used together with these density measurements to estimate $\gamma_{\rm cw}$.

\subsection{Measuring bulk density fluctuations}
\label{sec:alp}

Within the probabilistic model of \eqref{eqn:thp}, the parameters $\alpha_{l,v}$ are related to density fluctuations in the liquid and vapor phases.  Consider a homogeneous state of one phase, and separate the system into two sub-volumes: a box of size $V_b$ and the remainder of size $V-V_b$.  Writing $N_b$ for the number of particles in the box, the analogue of \eqref{eqn:thp} is
\begin{equation}
    P(N_b) \propto \exp \left[-V_b f \left(\frac{N_b}{V_b}\right) - (V-V_b) f \left(\frac{\rho V-N_b}{V-V_b} \right)  \right] \; .
\label{eqn:bulk-thp}
\end{equation}
where $f=f_l$ or $f=f_v$ depending on the phase considered, and $\rho=N/V$ is the total density.  Assuming that the box $V_b$ and the remainder $V-V_b$ are both large, the densities in both regions remain close to $\rho$.  Hence one may Taylor expand the functions $f$ about $\rho$ to obtain
\begin{equation}
P(N_b) \propto \exp\left[ -\alpha V\frac{(N_b-\rho V_b)^2}{2 V_b(V-V_b)} \right] \; ,
\end{equation}
with $\alpha=f''(\rho)$.  This is a Gaussian distribution with mean $\rho V_b$ and variance

\begin{equation}
    \frac{\mathrm{Var} (N_{b})}{V_{b}} = \frac{1}{\alpha} \left( 1 - \frac{V_{b}}{V} \right) \; .
    \label{eqn:alpha}
\end{equation}
This amounts to a prediction for $\mathrm{Var} (N_{b})/V_{b}$, which is valid when both $V_b$ and $V-V_b$ are large enough that fluctuations in both sub-volumes are Gaussian.
The same result has been derived previously in several other contexts~\cite{roman1999fluctuations,roman1997fluctuations, villamaina2014thinking, heidari2018fluctuations,vrins2018sampling}.

For a homogeneous equilibrium system with density $\rho$, $1 / \alpha = k_B T \rho^2 \kappa$, where $\kappa$ is the isothermal compressibility of the system.  For an ideal gas configuration (off-lattice) this means $\alpha = 1/\rho$.  For a \emph{lattice} ideal gas 
\begin{equation}
\alpha^{\rm ig} = \frac{1}{ \rho(1-\rho) } \; .
\label{equ:alpha-ig}
\end{equation}

A reliable method for estimating $\alpha$ from simulation data is then to measure and plot $\mathrm{Var}(N_{b})/V_{b}$ against $V_b$ (see Appendix~\ref{app:bulk} for an example in equilibrium). If one prepares a homogeneous phase at the coexistence density $\rho_{v,l}$, one may then identify $\alpha$ with $\alpha_{v,l}$ from \eqref{eqn:alpha}. We note that this assumes that the $f$ describing bulk fluctuations of the liquid in \eqref{eqn:bulk-thp} is the same one that describes phase coexistence in \eqref{eqn:thp}: we return to this point in Sec.~\ref{sec:bulk}.

\subsection{Generic theory for active interfaces}
\label{sec:flux}

The probabilistic arguments given so far rely on strong assumptions (\ref{eqn:thp},\ref{eqn:bulk-thp}) about steady-state probability distributions in large systems.  We will see that the ALG is not fully consistent with these theories.  For this reason, we give short theoretical arguments as to what may be expected generically for persistent active systems at phase coexistence.

Comparing slab and droplet configurations, one expects the densities to be corrected by curvature-dependent effects, as modeled in \eqref{eqn:lap}.  Expanding \cite{lee2017interface} in powers of $1/R$ (up to second order) one obtains
\begin{equation}
    \rho_v^d = \rho_v^s \left(1 + \frac{\xi_v}{R} - \frac{\xi_v\delta_v}{R^2} \right) \; ,
    \label{eqn:fb}
\end{equation}
for the vapor, 
and similarly
\begin{equation}
    \rho_l^d = \rho_l^s \left(1 + \frac{\xi_l}{R} - \frac{\xi_l\delta_l}{R^2} \right) \; ,
    \label{eqn:fb-l}
\end{equation}
for the liquid.  In equilibrium, \eqref{eqn:lap} gives $\xi$ in terms of the surface tension and the compressibility.  Moreover $\xi_v / \xi_l = \rho_l^s \alpha_l/\rho_v^s \alpha_v$ due to \eqref{eqn:lapcp}.  Also, $\delta_l,\delta_v$ would both be identified in equilibrium with the Tolman length~\cite{tolman1949effect}, which can be interpreted in two ways: (i) as a curvature-dependent correction to the surface tension, or (ii) as the distance between the equimolar dividing surface (the surface of the liquid determined by assuming a sharp boundary, using the lever rule) and the surface of tension.  When fitting data for the ALG  we show below that $\delta$ can be significant, which we attribute to significant interfacial thickness (see below).

Given the analogy with equilibrium, we define effective surface tensions associated with the density differences between slab and droplet.  For the vapor, we measure both $\xi_v$ and the bulk $\alpha_v$ and we define
\begin{equation}
   \gamma_v^{\rm eff}  =  \xi_v\alpha_v  \rho_v^s (\rho_l^s - \rho_v^s) , \quad 
    \label{eqn:tg}
\end{equation}
by analogy with \eqref{eqn:lap}.  A corresponding $\gamma_l^{\rm eff}$ can also be written down by substituting in \eqref{eqn:lapcp}.  In equilibrium
$\gamma_l^{\rm eff}= \gamma_v^{\rm eff}=\gamma_{\rm cw}$ but none of these equalities need to hold in general.

One way to understand \eqref{eqn:fb} is to consider the the Kelvin effect, where liquid-vapor interfaces that are curved away from the liquid phase have elevated vapor pressure \cite{thomson1871lx}. In other words, particles on these curved interfaces have higher rates of detachment than from a flat one. A kinetic argument that incorporates this effect has been derived \cite{redner2013Jul, redner2016classical, lee2017interface}.

\section{Phase separated states of the ALG}
\label{sec:algres}

\subsection{Overview of phase behavior}

We present numerical simulations of the ALG.  All results are for the steady state of the system, the transient relaxation into this state is discussed in Sec.~\ref{sec:dyn}.  To initialize the system, 
we place particles in either a uniform slab or a square droplet with volume fraction $0.9$ (this volume fraction is chosen to be slightly lower than the steady state liquid density).    For the slab, particle orientations in the outer third point inwards, with the others being random.   For the droplet, all orientations point towards the center.  With this choice, the system settles into the non-equilibrium steady states shown in Fig.~\ref{fig:mips}.  .

\begin{figure}
    \centering
    \includegraphics[width=1.0\linewidth]{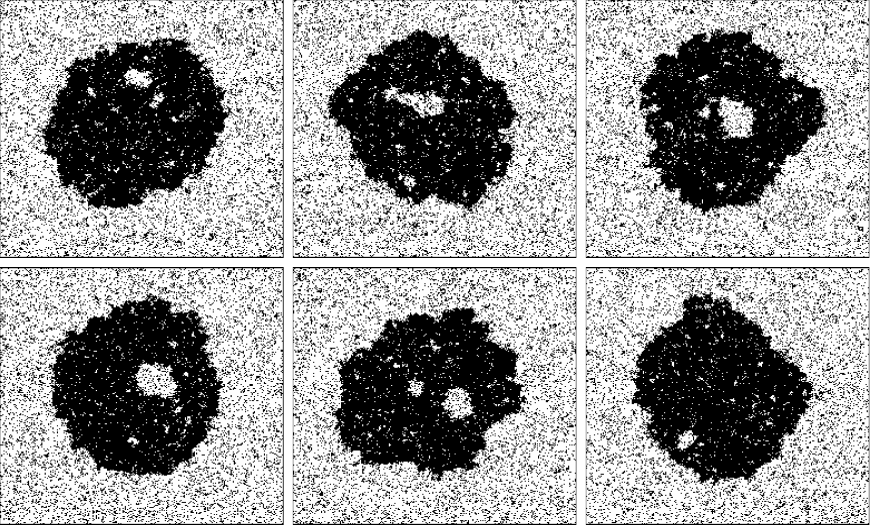}
    \caption{Time-ordered snapshots of the ALG in the droplet geometry at P1: $v_0 = 50$, $D_O = 1$, $D_E = 10$, $\phi = 0.4$ and $L = 256$, showing the formation and expulsion of a large bubble.}
    \label{fig:dropts}
\end{figure}

Fig.~\ref{fig:mips} shows that while the vapor region is roughly homogeneous, the liquid region is distinctively bubbly for P1 and P2. This is in agreement with the bubbly phase separation predicted by the reverse Ostwald ripening picture \cite{tjhung2018cluster}, and previous research on similar models \cite{stenhammar2014phase, shi2020selforganized, fausti2024statistical}. Remarkably, some large bubbles in the dense region appear to grow indefinitely until their expulsion into the bulk vapor.  This is illustrated in Fig.~\ref{fig:dropts}. We will discuss this phenomenon in greater detail in Sec.~\ref{sec:dyn}. 

For parameters P3, we do not find bubbles in systems with $L = 128$ to $256$, these systems look similar to panel P3 of Fig.~\ref{fig:mips}.  This panel shows a typical configuration at $L = 512$ which is also not bubbly.  However, we observe that (i) transient bubbles form at the beginning of the trajectory, where the liquid density is prepared to be lower than its steady state value, (ii) these bubbles are stable until they are expelled into the bulk vapor, and (iii) the nucleation and expulsion of small bubbles very occasionally happen in the steady state. In this sense, the lack of bubbles for P3 is likely to be a finite-size effect, and we expect more and larger bubbles to form in larger system sizes.

The three state points in Fig.~\ref{fig:mips} have increasing $D_E$, which amounts to an increase of thermal noise.  One sees that this makes the interfaces less sharply-defined.  It is useful then to consider the ``interfacial thickness'' which is the length scale over which the density changes from $\rho_l$ to $\rho_v$ (this is distinct from the interfacial width $w$ defined in \eqref{eqn:cwtw} which instead measures fluctuations in the position of the interface).  The fact that the very small bubbles visible in P1 are suppressed in P2 appears to be linked to the larger interfacial thickness, which sets a lower cutoff scale below which bubbles are not stable.  Note that the hydrodynamic limit of Ref.~\onlinecite{mason2023exact} results in an interfacial thickness that is much larger than the lattice spacing $a$, which suppresses bubbles completely.

\subsection{Densities of coexisting phases}
\label{sec:rho}

\begin{figure}
    \centering
    \includegraphics[width=1.0\linewidth]{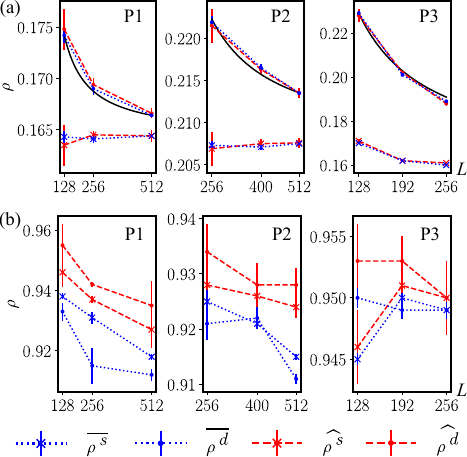}
    \caption{Mean and mode of the measured densities in the slab and droplet geometries for the ALG at (from left to right for each sub-figure) $v_0 = 50$, $D_E = 10$, $\phi = 0.4$; $v_0 = 50$, $D_E = 50$, $\phi = 0.4$; and $v_0 = 110$, $D_E = 200$, $\phi = 0.45$, for (a) the vapor and (b) the liquid. The solid lines are fits of \eqref{eqn:fb}.}
    \label{fig:rfit}
\end{figure}

Within the phase-separated steady states of the system, we collect histograms of the density within the coexisting phases, using sampling boxes of size $V_b$. The histograms are shown in Appendix~\ref{sec:hist} for various system sizes, along with the box sizes used. For systems undergoing bubbly phase separation, we treat the bubbles as part of the liquid phase: they reduce the mean density and to skew the distribution to the left.

The mean densities, denoted by $\overline{\rho}$, of the liquid and vapor phases, are shown in Fig.~\ref{fig:rfit}.  Since the total density is held constant, the radius of the liquid domain in the droplet geometry grows proportional to $L$.  Hence the interfacial curvature is smaller in large systems, and one expects the densities in slab and droplet geometries to approach each other as $L\to\infty$.  For the low-density (vapor) phase (Fig.~\ref{fig:rfit}(a)), the mean density in the slab geometry is almost independent of system size, while the density in the droplet geometry decreases with $L$.  This is consistent with the prediction of \eqref{eqn:fb}, the fitting parameters are shown in Table~\ref{table:data}. The Tolman length $\delta$ in \eqref{eqn:fb} cannot be neglected for our systems, due to the thickness (or the lack of sharpness) of the interfaces: for interfaces whose density profiles vary over an extended lengthscale, the equimolar dividing surface is unlikely to be a good approximation of the surface of tension.

We also computed the most likely values of the density (the modes of the distributions), which are denoted $\widehat{\rho}$.  The mean and mode are close in all cases for the vapor phase, and the histograms appear symmetric. This is as expected since the vapor appears homogeneous in Fig.~\ref{fig:mips}.

\begin{figure}
    \centering
    \includegraphics[width=1.0\linewidth]{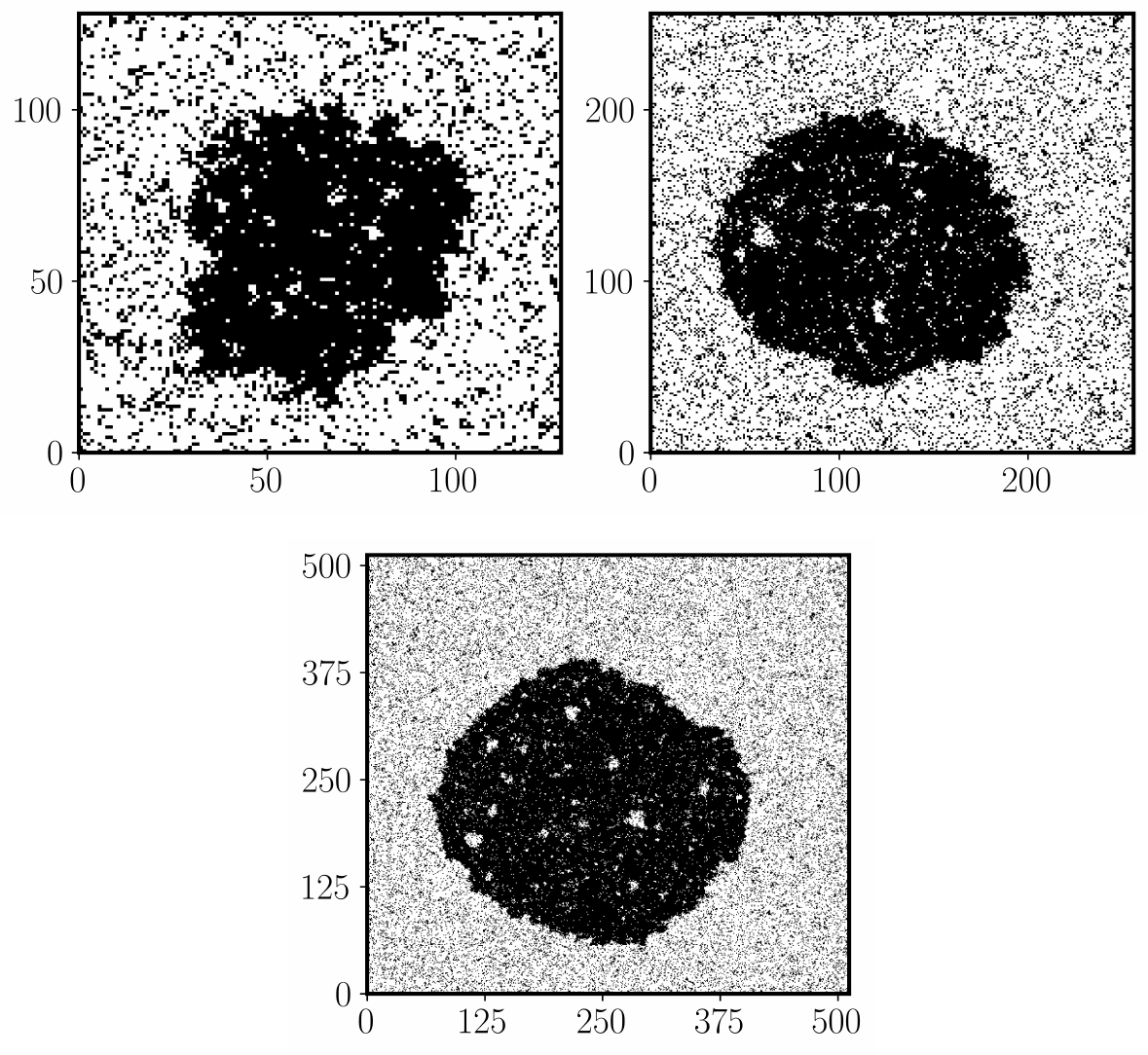}
    \caption{Typical configurations of the ALG in the droplet geometry at P1: $v_0 = 50$, $D_O = 1$, $D_E = 10$, $\phi = 0.4$ for $L = 128$, $256$ and $512$.}
    \label{fig:sizes}
\end{figure}

Turning to the liquid phase, the situation is more complex.
For P1 and P2, where bubbly phase separation is observed, the mean densities obey $\overline{\rho_l^d} < \overline{\rho_l^s}$ (similar to the droplet), but the most likely (mode) densities have the opposite ordering $\widehat{\rho_l^d} > \widehat{\rho_l^s}$. This arises because the distributions are skewed, with tails at low density due to the bubbles inside the liquid (as clearly shown in Appendix~\ref{app:rhosd} by Fig.~\ref{fig:rho}). The skew is more pronounced for the droplet geometry, which we attribute to the fact that the diameter of the droplet is larger than the thickness of the corresponding slab, allowing larger bubbles to develop.

The averages $\overline{\rho_l^s}$ and $\overline{\rho_l^d}$ in this regime both decrease weakly with increasing $L$. 
This can be explained by considering the bubble size distribution: the minimum size of stable bubbles is determined by the interfacial thickness, independent of the system size. Bubbles also have a typical lifetime, since they are eventually expelled into the vapor: this gives an upper cutoff to their size distribution which grows with the linear size of the liquid cluster, which in turn grows with the system size.
This is illustrated in Fig.~\ref{fig:sizes}, where typical droplet clusters for P1 are plotted for various system sizes. The key observation is that the sizes of the largest bubbles in these snapshots grow with the system size (note the scale of these plots). As a consequence, larger systems contain more large bubbles in the liquid, causing the measured liquid density to decrease with system size, as shown in Fig.~\ref{fig:rfit}. 

For P3, we consider $L\leq 256$ where bubbly phase separation is not observed.
The mean and mode of the liquid density for these parameters are separately very similar between the slab and droplet geometries (see Fig.~\ref{fig:rfit}(b) and note the range of the $y$ axis for P3).  At the largest size all four measurements are very close, within numerical error range. In an equilibrium system, this might be expected from \eqref{eqn:lapcp} because liquids tend to have low compressibility compared to the vapor, corresponding to small density fluctuations and large $\alpha$. However, we will see in Sec.~\ref{sec:bulk} that these density fluctuations are not small in the ALG.

\subsection{Bulk density fluctuations}
\label{sec:bulk}

\begin{table}
\centering
\begin{tabular}{| c | c | c | c |} 
    \hline
    Parameters & $\alpha_v$ & $\tilde{\gamma}$ & $\delta$ \\ 
    \hline
    P1 & $3.28 \pm 0.07$ & $1.99 \pm 0.06$ & $7.3 \pm 2.3$ \\ 
    \hline
    P2 & $2.75 \pm 0.03$ & $3.72 \pm 0.52$ & $30 \pm 22$ \\ 
    \hline
    P3 & $3.94 \pm 0.05$ & $13.7 \pm 0.18$ & $10.3 \pm 0.8$ \\ 
    \hline
\end{tabular}
\caption{The values of the fitted quantities $\alpha_v$, $\tilde{\gamma}$ and $\delta$. $\alpha_v$ are extracted from Fig.~\ref{fig:bulk}, while $\tilde{\gamma}$ and $\delta$ are extracted from Fig.~\ref{fig:rfit}. For comparison, we estimate $\alpha_l= 0.37 \pm 0.01$ for P1.}
\label{table:data}
\end{table}

We now measure the density fluctuations in bulk simulations of the active liquid and vapor states to determine the quantity $\alpha$ required in \eqref{eqn:lapcp} and \eqref{eqn:lap}. We determine the densities to be used in these bulk simulations by taking the average densities in the slab geometries shown in Fig.~\ref{fig:rfit}. While the densities of the vapor are relatively stable, densities in the liquid phase decrease weakly with system size for parameters P1,P2. In this case we use the average densities obtained from the second largest system sizes for our bulk simulations (see further discussion below).  

\begin{figure}
    \centering
    \includegraphics[width=1.0\linewidth]{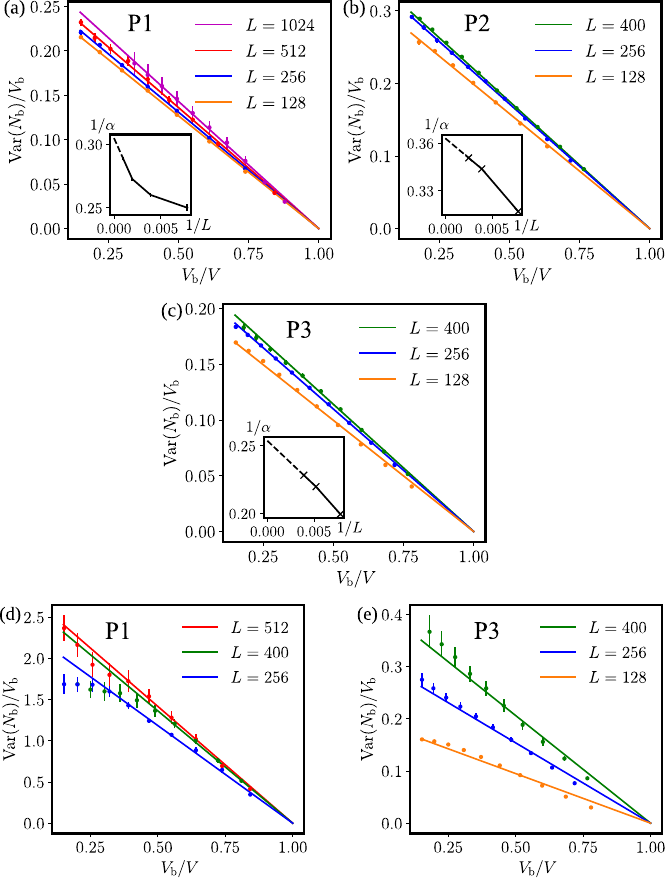}
    \caption{Bulk density fluctuations of the ALG. (a,b,c) Vapor phase.  Parameters are P1,P2,P3 as shown, except that the densities are $\phi=(0.164,0.207,0.160)$ for the three panels, chosen to match the vapor phase densities of Fig.~\ref{fig:rfit}.
    (d,e) Liquid phases for P1,P3 with $\phi=(0.931,0.950)$.  Insets show $1 / \alpha$ against $1 / L$. The error bars of the data points in the insets are omitted as they are smaller than the symbol size. The dashed lines are linear extrapolations.  Measurements of the liquid density fluctuations for P2 are not shown: these observables require a large computational time to obtain converged results.
    }
    \label{fig:bulk}
\end{figure}

For these calculations, we initialize homogeneous systems at the chosen densities, and allow them to reach steady states, before measuring density fluctuations in boxes of various sizes $V_b$ (recall \eqref{eqn:alpha}).
The data and its fit to \eqref{eqn:alpha} are shown in Fig.~\ref{fig:bulk}, where the intercept with the vertical axis gives the value of $1/\alpha$.  For the vapor phase, {fluctuations appear Gaussian, and the data is well fitted by \eqref{eqn:alpha} for each system size.} However, the inferred value of $1 / \alpha$ drifts upwards for larger systems, and has not fully converged within accessible system sizes. A similar effect is also seen in equilibrium (see Appendix \ref{app:bulk}). {We attribute this to an effect of periodic boundaries, where finite correlation lengths cause an undercounting of fluctuations that are correlated across the periodic boundaries.} Therefore, the finite-size correction to $1 / \alpha$ should be proportional to the surface-to-volume ratio $1 / L$, so we extrapolate to $L\to\infty$ by plotting $1/\alpha$ against $1 / L$, as seen in the insets of Fig.~\ref{fig:bulk}. {Note from the $y$-axis of these insets that the measured value of $\alpha$ is not expected to be significantly dependent on the exact extrapolation scheme.} The values of $\alpha_v$ are reported in Table~\ref{table:data}.  For lattice ideal gases at the same densities, \eqref{equ:alpha-ig} yields $\alpha_v^{\rm ig} = (7.29, 6.09, 7.44)$ for P1, P2 and P3 respectively: the smaller values in Table~\ref{table:data} are indicative of clustering within the vapor.
{The question of extracting $\alpha$ from equilibrium simulations has been re-examined recently, either by exploiting the grand canonical ensemble, or by integrating the two-point density correlation function up to a cutoff (instead of measuring $N_b$ in a finite box)~\cite{wilding2024what}. It would be useful to assess the various methods in nonequilibrium systems too.}

Due to the large bubbles in the liquid phase, density fluctuations in the bubbly liquid (P1 and P2) are much larger than in the vapor and in liquids without bubbles (P3). Fig.~\ref{fig:bulk} shows results for P1 and P3 (results for P2 are not shown since converging these numerical calculations require long simulations of large systems and P1,P3 are sufficient to illustrate the behavior). The fluctuations are much larger in the bubbly case (P1). The bubbles also contribute to correlations over large length scales, leading to {non-Gaussian fluctuations} and deviations from \eqref{eqn:alpha} at small $V_b$.

\begin{figure}
    \centering
    \includegraphics[width=1.0\linewidth]{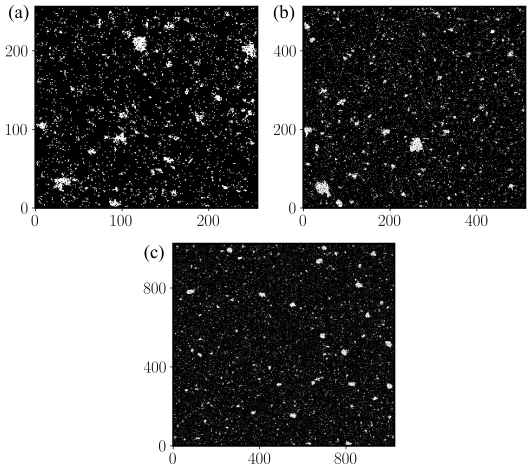}
    \caption{Snapshot of the bulk liquid at  $\phi = 0.931$ for the ALG at $v_0 = 50$, $D_O = 1$, $D_E = 10$ (P1) and (a) $L = 256$, (b) $L = 512$ and (c) $L = 1024$. Note that bubbles here do not grow with system size.}
    \label{fig:liq}
\end{figure}

Fig.~\ref{fig:liq} shows snapshots of the bubbly liquid. The typical size of bubbles grows more slowly than the system size, so that the system looks increasingly homogeneous on large scales.  This is the expected behavior for microphase separation, {where there exists a characteristic domain size that is independent of system size. This stands in contrast to finite-size scaling behavior at phase coexistence, where bubbles inside the slab (or droplet) configurations grow with system size (recall Fig.~\ref{fig:sizes}).}

The existence of mesoscopic bubbles is in agreement with the reverse Ostwald ripening picture \cite{tjhung2018cluster}, where small bubbles grow at the expense of larger ones.
One way to understand this process in the ALG  is to consider the transport of vacancies into and out of bubbles.  Reverse Ostwald ripening can occur if the flux of vacancies away from a larger bubble tends to be larger: that is ``bubble detachment'' is slower for more positively curved interfaces.  It would be interesting to identify a specific physical mechanism for this differential flux, which might be possible in models of this type.

For P3, where bubbles are absent, we observe much smaller density fluctuations (compare the $y$ axes of Figs.~\ref{fig:bulk}(d) and (e)). {However, density histograms remain non-Gaussian, with skewness towards lower densities at small $V_b$, and skewness towards higher densities at large $V_b$, leading to systematic deviations from \eqref{eqn:alpha}.}

Within the probabilistic theory of Sec.~\ref{sec:prob}, these measurements of $\alpha$ in bulk phases can be related to systems at phase coexistence, including the results shown in Fig.~\ref{fig:rfit}. This will be discussed further below, but we can already notice here that \eqref{eqn:lapcp} does \emph{not} hold in our system.  Specifically, \eqref{eqn:lapcp} predicts
\begin{equation}
\alpha_l = \frac{\alpha_v  (\rho_v^d - \rho_v^s) }{ \rho_l^d - \rho_l^s } \; .
\label{eqn:alpha-l-predict}
\end{equation}
That is, the ratio of the density fluctuations for the two phases (quantified by $\alpha$) is equal to the ratio of the density shifts between droplets and slabs, as anticipated in Sec.~\ref{sec:intro}.
Noting that $\alpha_{v,l}$ are strictly positive by definition, \eqref{eqn:alpha-l-predict} requires that the density differences $\rho_v^d - \rho_v^s$ and $\rho_l^d - \rho_l^s$ have the same sign. For the bubbly liquids P1 and P2,  Fig.~\ref{fig:rfit} shows that these density differences have opposite signs, in direct contradiction to this relation. For P3, where no bubbles are observed for the system sizes chosen, we measure $\rho_l^d - \rho_l^s = 0.000 \pm 0.003$, such that \eqref{eqn:alpha-l-predict} requires a very large $\alpha_l$, while we measure $\alpha_l$ to be around $1.7$. Therefore, we conclude that either the probabilistic model \eqref{eqn:thp} is simply not applicable, or the density fluctuations of the bulk phase are so different to those at phase coexistence that $\alpha_l$ in cannot \eqref{eqn:thp} be related to the corresponding bulk $\mathrm{Var}(N_b)$ of \eqref{eqn:alpha}. 

The latter scenario is expected, as we have already shown that the finite-size scaling of liquid density fluctuations are qualitatively different between the bulk and phase-coexistence states (compare Figs.~\ref{fig:sizes} and \ref{fig:liq}). This difference can be rationalized by a general lack of ensemble equivalence in nonequilibrium systems: that is, the behavior of the liquid inside a phase separated droplet may be different from that inside a slab, and both may differ from a homogeneous bulk liquid at the same density. Such differences are generic signatures of long-ranged (non-local) interactions, in violation of \eqref{eqn:thp}. In the specific case considered here,  the vapor in the phase-separated state can be seen as a bath for vacancies which is absent in the bulk liquid systems. 

\subsection{Capillary waves}
\label{sec:algcwt}

This Section analyzes 
the interfacial fluctuations in of the ALG, and compares them to CWT. Specifically, we measure the spectrum $\langle |h_q | ^2 \rangle$ as a function of $q$ and we compare its low-$q$ behavior with the corresponding prediction \eqref{eqn:cwt} of CWT.  
The interfacial positions $h_y$ are determined by detecting the largest connected cluster of occupied sites in the system, and finding its extrema in the $x$ direction for each value of $y$ (therefore the outer edges of any overhang are registered as the interface). This procedure yields two $h_y$ measurements per snapshot (one for each interface).   
 
\begin{figure}
    \centering
    \includegraphics[width=1.0\linewidth]{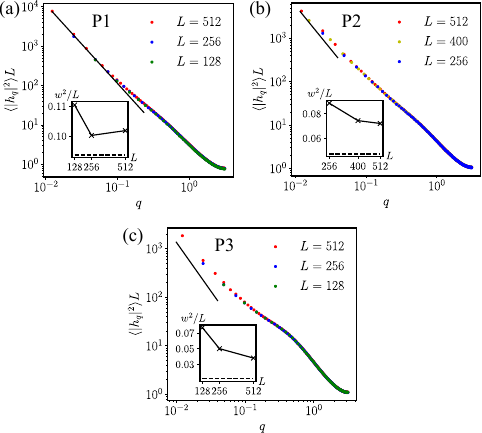}
    \caption{The interfacial spectrum of the ALG, at (a) $v_0 = 50$, $D_E = 10$, $\phi = 0.4$, (b) $v_0 = 50$, $D_E = 50$, $\phi = 0.4$, and (c) $v_0 = 110$, $D_E = 200$, $\phi = 0.45$. The straight lines have gradient $-2$, and their intercepts are given by the surface tension extracted through \eqref{eqn:fb} and \eqref{eqn:tg}. Insets: measured interfacial width plotted against system size; the dashed lines are predictions from \eqref{eqn:wid}.}
    \label{fig:alg_cwt}
\end{figure}

Results are shown in
Fig.~\ref{fig:alg_cwt}, for several system sizes. The finite-size scaling is consistent with \eqref{eqn:cwt}.
We also plot solid lines that are proportional to $q^{-2}$, as required by CWT, see below for further discussion. For parameters P1, the system obeys CWT on large scales (small $q$).  For larger $q$, the fluctuations are larger than one obtains by extrapolating the small-$q$ behavior, indicating fluctuations larger than CWT predicts.

For P2,P3, the excess fluctuations at large $q$ persist throughout the of accessible length scales, although the data show strong signs of converging towards $q^{-2}$ scaling at small $q$. This effect also means that the interfacial width $w^2$ defined in \eqref{eqn:cwtw} exceeds the value estimated via \eqref{eqn:wid} for these finite systems (see the insets in Fig.~\ref{fig:alg_cwt}).  This motivates against estimating $\gamma_{\rm cw}$ via the interfacial width, which will result in an underestimation. Instead, one should extract $\gamma_{\rm cw}$ through a fit of the spectrum of $h_q$, where one can identify the range of $q$ where CWT holds.

The excess of $\langle |h_q | ^2 \rangle$ above the small-$q$ CWT scaling is present for P3 as well as P1,P2 so we do not attribute it to bubbles (see also Sec.~\ref{sec:dyn}).  It is also absent in the passive system of Appendix~\ref{app:cwt} so it is not intrinsic to the lattice or the surface detection algorithm.  It must therefore be attributable to some small-scale interfacial fluctuation mechanism of the active particles, and may be related to the interfacial thickness, which increases from P1 to P2 to P3.
It is interesting that the interfacial thickness has the opposite trend with parameters than the interfacial width $w$ (contrary to what is expected at equilibrium \cite{onsager1944crystal, fisher1984curvature}).

Finally, we discuss the the solid lines which indicate $q^{-2}$ scaling in the main panels of Fig.~\ref{fig:alg_cwt}. 
These lines are not fitted, they show
\begin{equation}
L \langle | h_q |^2 \rangle = \frac{1}{q^2 \gamma_v^{\rm eff} } \; ,
\label{equ:cwt-xi}
\end{equation}
where $\gamma_v^{\rm eff}$ is given by \eqref{eqn:tg}, obtained from the measurements in Figs.~\ref{fig:rfit} and \ref{fig:bulk}.  
These measurements do not use any properties of the interfaces, but \eqref{equ:cwt-xi} still shows very reasonable agreement with capillary wave spectrum at small $q$. This agreement is clearest for parameters P1. For the values of $q$ available to our simulations, the data for parameters P2 and P3 does not yet show full convergence to the small-$q$ scaling regime, so one may find better agreement in that regime.

To understand the physical interpretation of \eqref{equ:cwt-xi}, we rearrange it [using also (\ref{eqn:fb},\ref{eqn:tg})] to obtain
\begin{equation}
R(\rho^d_v - \rho^s_v) \approx \frac{1}{Lq^2 \langle |h_q|^2 \rangle \alpha_v (\rho^s_l - \rho^s_v ) } \; ,
\label{eqn:vap-shift-predict}
\end{equation}
where the approximate equality holds for small $q$ and large $R$. We interpret this formula as a prediction based on \eqref{eqn:thp}, which gives the curvature-induced density shift of the vapor phase in terms of the capillary wave surface tension and the bulk density fluctuations of the vapor.  (The right hand side can be computed using only slab and bulk simulations, without requiring the droplet geometry.)  In equilibrium, this reduces to the standard prediction of the density shift due to Laplace pressure~\cite{rowlinson1982molecular} (see also Appendix ~\ref{app:ising}).
Since Fig.~\ref{fig:alg_cwt} is consistent with \eqref{equ:cwt-xi}, we find that Eq.~\eqref{eqn:vap-shift-predict} also gives valid predictions for the density of the vapor between slab and droplet.

On the other hand, the theory of \eqref{eqn:thp} also predicts that  \eqref{equ:cwt-xi} should still hold on replacing $\gamma_v^{\rm eff}$ with $\gamma_l^{\rm eff}$: this prediction does not agree at all with the data, due to the strong violation of \eqref{eqn:alpha-l-predict} discussed in Sec.~\ref{sec:bulk} [this is equivalent to the violation of \eqref{eqn:lapcp}].
Similarly, using \eqref{equ:cwt-xi} with (\ref{eqn:fb-l},\ref{eqn:tg}) gives 
\begin{equation}
R(\rho^d_l - \rho^s_l) \approx \frac{1}{Lq^2 \langle |h_q|^2 \rangle \alpha_l (\rho^s_l - \rho^s_v ) } \; ,
\label{eqn:liq-shift-predict}
\end{equation}
which is a prediction of the density shift of the \emph{liquid}  based on \eqref{eqn:thp}.  This prediction also fails, for the same reason.

In summary, \eqref{eqn:vap-shift-predict} and \eqref{eqn:liq-shift-predict} are two predictions of the simple theory \eqref{eqn:thp}: one of them is consistent with the data while the other disagrees strongly. These two predictions are dependent on each other within the theory, so the numerical agreement of \eqref{eqn:vap-shift-predict} with our data is surprising. However, simulations suggest that the vapor in our model resembles a simple homogeneous fluid, so its equilibrium-like properties are not unexpected. We will suggest some rationalizations of \eqref{eqn:vap-shift-predict} in Sec.~\ref{sec:algcon}. On the other hand, the failure of \eqref{eqn:liq-shift-predict} means that some refinement of the theory in \eqref{eqn:thp} is needed to describe our model, as well as general models of active matter. In particular, it would be valuable to understand how \eqref{eqn:thp} might be modified to account for the existence of bubbles.

\subsection{Dynamics of bubble formation}
\label{sec:dyn}

In the phase-separated steady state, the bubbly liquids for P1 and P2 are constantly boiling, where bubbles go through the process of nucleation, growth, followed by expulsion into the vapor. In Fig.~\ref{fig:dropts}, we have shown this process in a series of snapshots evenly ordered in time for a system in the droplet geometry.  Here we analyze the phase-separated dynamics in more detail.

\begin{figure}
    \centering
    \includegraphics[width=0.8\linewidth]{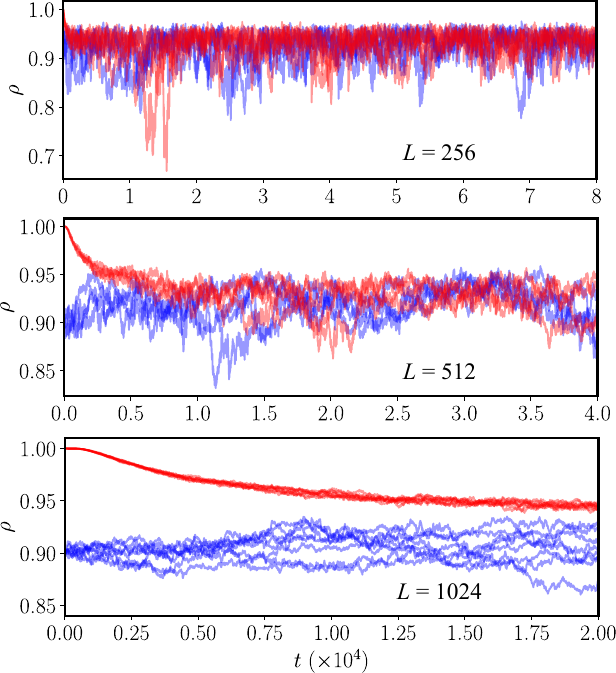}
    \caption{Time series of the liquid density in the phase separated ALG in the droplet configuration, at $v_0 = 50$, $D_O = 1$, $D_E = 10$ (P1) and $L = 256$ to $1024$. The red and blue curves are for slabs prepared with density $1$ and $0.9$, respectively. 4 trajectories for each initial condition are shown for the $L = 256$ and $512$ cases, while 8 are shown for the $L = 1024$ case.
    Note the total time simulated decreases with system size: the range shown is sufficient to illustrate the relevant behavior.
    }
    \label{fig:rhots}
\end{figure}

Fig.~\ref{fig:rhots} shows time series in droplet geometry for parameters P1 and two different initial conditions.  Specifically, we initialize a droplet either with density $\rho=1$ or by placing particles randomly in a square region at density $\rho=0.9$ (lower than the liquid density $\rho_l\approx 0.94$ from Fig.~\ref{fig:rfit}).  We plot the density inside the droplet. For system sizes $L=256,512$, one observes convergence within a simulation time $t=10^4$ (in units of $D_O$: recall Sec.~\ref{sec:algmod}) to a steady state that is independent of initial condition. The large fluctuations in this steady state are due to the bubbles, and the time scale for these fluctuations grows with system size. We note that the shape relaxation (roughening) of the droplet from square to (almost) circular is fast compared to these time scales.

For the largest system $L=1024$, the system does not converge within the time window $t=2\times 10^4$ shown. This is due to the long time scale associated with vacancy movement in the bulk of the slab, which, {assuming such movements are diffusive,} scales as the square of the droplet radius {(this $L^2$ dependence has been observed in AMB+ \cite{fausti2024statistical})}.

Together, these data illustrate the computational difficulty of finite-size scaling in these systems, especially given the large time scales associated with bubble formation and motion.  Hence, even if the system sizes in (for example) Fig~\ref{fig:bulk} and Fig.~\ref{fig:alg_cwt}(b,c) are not sufficient to fully converge to the large-$L$ limit, it is not straightforward to analyze yet larger systems.

On the other hand, for all system sizes the vapor density relaxes much faster than the liquid density. For example, for $L = 1024$, where the liquids prepared in either initial states have not converged to the steady state within our simulation time, which is of order $10^4$, we find that the vapors have already reached their steady states by a time of around $10^3$ (data not shown).

\begin{figure}
	\centering
	\includegraphics[width=1.0\linewidth]{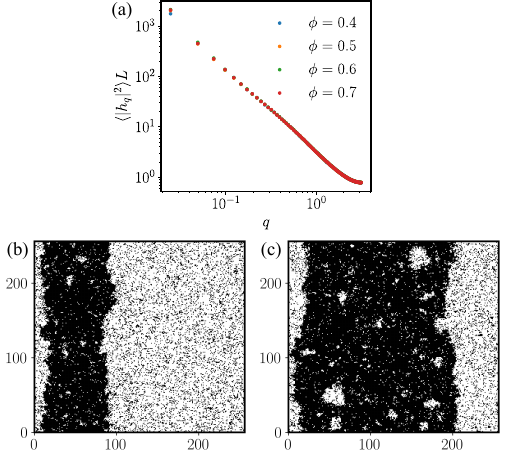}
	\caption{(a) The interfacial spectrum of the ALG with $L=256$ and parameters P1,  except that $\phi$ is varied from $0.4$ to $0.7$  (b) A typical slab configuration at $\phi = 0.4$. (c) Similar snapshot for $\phi = 0.7$}
    \label{fig:cwtbub}
\end{figure}

Finally, we consider possible effects of bubbles on the capillary wave spectrum.  In principle, the expulsion of bubbles from a slab or droplet acts as a source of interfacial fluctuation.  To test this, we prepared steady-state slab configurations with different total volume fractions ($\phi = 0.4$ to $0.7$), as shown in Fig.~\ref{fig:cwtbub}, where on sees that the bubble sizes are larger when the slabs are thicker. However, Fig.~\ref{fig:cwtbub}(a) shows that the CWT spectra are indistinguishable, indicating that the bubble size distribution does not affect the capillary waves. This is likely due to the bubble expulsion events being relatively rare within the timescale of interfacial fluctuations, so that the capillary waves quickly relax and heal any bubble-induced fluctuations. Considering that the vapor density relaxes much more quickly than the liquid density, the quick relaxation of capillary waves is most likely due to currents in the vapor phase from the parts of the interface with positive curvature to those with negative curvature. These currents have been observed in simulations of off-lattice models \cite{turci2024partial}, and can be considered a consequence of the curvature dependence of the vapor density shown in Fig.~\ref{fig:rfit}, in direct analogy to equilibrium Ostwald ripening.

\section{Discussion and outlook}
\label{sec:algcon}

{Let us summarize the main conclusions of our extensive simulations. Our ALG supports MIPS as well as bubbly phase separation (Fig.~\ref{fig:mips}). Sec.~\ref{sec:algcwt} showed that the fluctuations of the liquid-vapor interface are consistent with CWT for very small wavevectors $q$ (large length scales), but the fluctuations at larger wavevectors seem to exceed CWT predictions. For droplet geometries, the coexisting phases have density shifts that are functions of the droplet radii (Fig.~\ref{fig:rfit}). For the dilute (vapor) phase, the density of the droplet geometry exceeds that of the slab geometry. We explained in Sec.~\ref{sec:algcwt} that our results for this density shift show reasonable agreement with a prediction of the theory \eqref{eqn:thp}, based on capillary wave fluctuations and bulk density fluctuations of the droplet. In other words, the same effective surface tensions governs capillary waves and the curvature dependence of the vapor density around curved interfaces. In equilibrium, this is expected, and can be quantified by the effects of Laplace pressure. On the other hand, the observed density shift of the liquid phase cannot be predicted by the theory. We also analyzed density fluctuations within homogeneous liquid and vapor phases (Sec.~\ref{sec:bulk}).

We stress that the predictions of \eqref{eqn:thp} for the vapor and liquid phases are not independent: the failure of this theory for the liquid phase should in principle render it equally inapplicable to the vapor. Therefore, the apparent agreement between the theory and data for the vapor is surprising. However, this agreement is not entirely unjustified. First, there is a separation of timescales between density relaxation of the vapor and the liquid, where the much faster observed relaxation of the vapor density implies that capillary wave fluctuations relax predominantly through currents in the vapor. Moreover, as the vapor in our system behaves like  a simple homogeneous fluid, the correct description of its fluctuations should be similar to the form of \eqref{eqn:thp}. Therefore, it would be interesting to check if our conclusion is robust in other models of MIPS, or if further insights can be extracted from dynamical hydrodynamic models such as AMB+.}

On the other hand, we note that no local model can describe bubble formation or the apparent violations of ensemble-equivalence between bulk liquids, slabs, and droplets.  The presence of non-local effective interactions can cause significant (fluctuating) particle currents in the steady state~\cite{spohn1983long, derrida2007non}.  The construction of such a non-local theory remains a challenge.

The tendency of the ALG towards bubbly phase separation depends on the underlying translational diffusivity $D_E$, with larger $D_E$ leading to thicker interfaces (crossing over more smoothly from liquid to vapor) and a suppression of small bubbles.  For parameters P3 we did not observe bubbles in the system sizes considered here, although we explained that these might reappear in larger systems.  The bubbly phase separated states appears consistent with previous work~\cite{tjhung2018cluster,shi2020selforganized,fausti2024statistical} in that bubbles are nucleated in the bulk of the liquid, and diffuse towards the interface, where they are expelled into the vapor.

The hydrodynamic limit of this ALG was analyzed in Ref.~\onlinecite{mason2023exact}: that limit corresponds in our notation to a very large $D_E \sim N$. Specifically, it appears that the large interfacial thickness in that regime tends to suppress bubble formation. This conclusion, and the Laplace-pressure-like curvature dependence of the excess vapor density around a curved interface, are expected to be general for systems undergoing MIPS.

Simulations of off-lattice models have also shown that the MIPS cluster is in fact a highly heterogeneous mosaic of finite-sized crystalline or hexatic domains, between which topological defects and bubbles can form \cite{redner2013structure, caporusso2020motility}. It is unclear whether this behavior can be readily captured by lattice models or hydrodynamic theories, but the connection between bubble formation and the geometric structure of the liquid merits further investigation.

Finally, recent work \cite{shi2020selforganized, fausti2024statistical} suggested that the bubble size distribution, instead of density, should be the central object in describing a liquid undergoing bubbly phase separation. However, the exact distinction between bubbles and liquid density fluctuations is difficult to establish numerically. 
In particular, investigating the dependence of the measured bubble size distribution on the coarse-graining scheme and the density cutoff can be helpful.

\begin{acknowledgments}
We thank Cesare Nardini for helpful discussions and a critical reading of the manuscript. We also thank Mike Cates and Filippo de Luca for helpful discussions.
\end{acknowledgments}

\appendix

\section{Interfacial and density statistics of the $2d$ Ising model}
\label{app:ising}

For demonstration, we will apply the theories described in Sec.~\ref{sec:algtheory} to a $2d$ Ising model with conserved (Kawasaki) dynamics, where the total number of up and down spins are fixed. The elementary proposed moves of this dynamics are the exchange of a particle with one of its nearest neighbors, which is then accepted or rejected with Metropolis rate. We identify the up spins as particles and the down spins as vacancies, and choose temperature $T = 2.0$, sufficiently below $T_c$ to avoid critical fluctuations. We then use the same notations as in our ALG, where the system size is $V = L^2$, and the overall volume fraction is $\phi = 0.4$. The algorithms used to conduct the measurements reported in this Appendix are identical to those used for the ALG.

\subsection{Capillary waves}
\label{app:cwt}

\begin{figure}[b]
    \centering
    \includegraphics[width=0.7\linewidth]{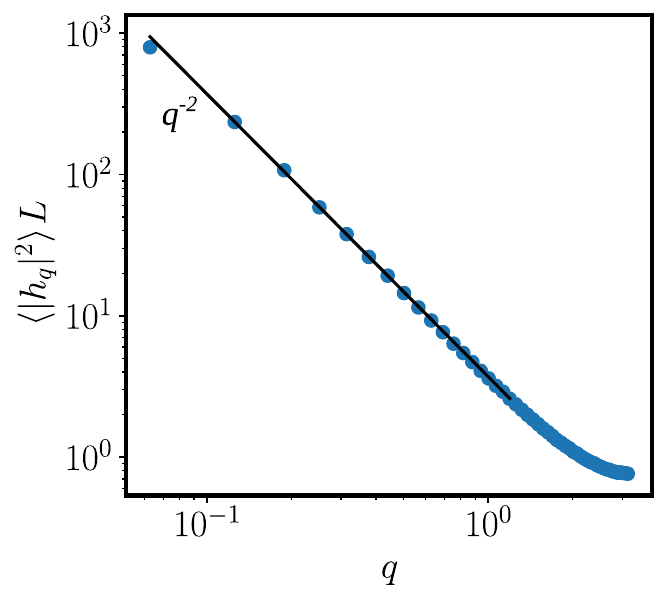}
    \caption{The interfacial spectrum of the $2d$ Ising model, at $T = 2.0$, $\phi = 0.4$ and $L = 100$. The straight line has gradient $-2$.}
    \label{fig:pas_cwt}
\end{figure}

Fig.~\ref{fig:pas_cwt} shows that in this passive system CWT is well satisfied: on a log-log scale, a straight line of gradient $-2$ fits well up to $q$ values of order $1$, beyond which the the microscopic details of the interface become significant. From the intercept of this plot we can extract a value of the surface tension $\gamma = 0.268 \pm 0.008$ for our choice of parameters, which is close to the value $0.23$ predicted by the exact solution of this model \cite{onsager1944crystal}. The error bar of $\gamma$ is the standard error of $\gamma$ measured on independent CPU threads. The slight discrepancy between the measured and predicted values may be due to finite-size effects, as the systems size used in our measurements ($L = 100$) is quite small.

\subsection{Densities in the phase-separated state}
\label{app:rhosd}

\begin{figure}
    \centering
    \includegraphics[width=1.0\linewidth]{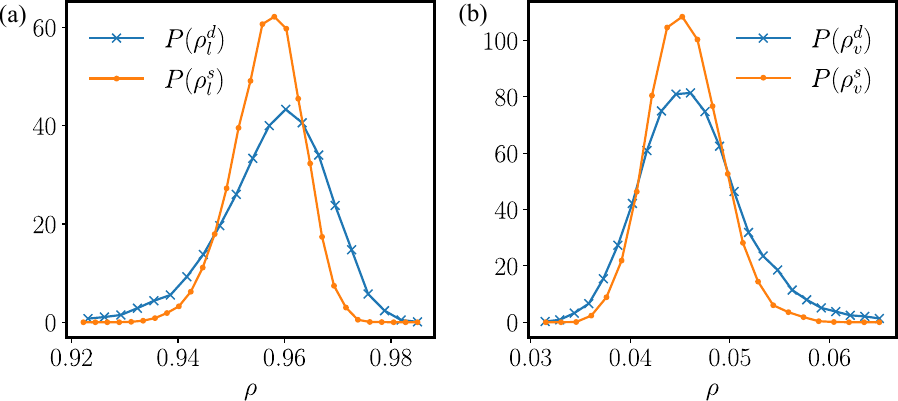}
    \caption{Histograms of densities measured within the (a) liquid and (b) vapor phase of the $2d$ Ising model at $T = 2.0$, $\phi = 0.4$ and $L = 200$. Note that the widths of the distributions are influenced by the box sizes $V_b$, which are 16000, 7600, 4000 and 1600, for measurements of $\rho_v^s$, $\rho_v^d$, $\rho_l^s$ and $\rho_l^d$, respectively.}
    \label{fig:pas_den}
\end{figure}

We show the density histograms of the phase-separated Ising system in Fig.~\ref{fig:pas_den}. The average densities are slightly higher in the droplet geometry than in the slab geometry, as predicted by the theory in Sec.~\ref{sec:alp}.  (The widths of these histograms are dependent on $V_b$, whose values are given in the caption.)

\subsection{Bulk density fluctuations}
\label{app:bulk}

\begin{figure}
    \centering
    \includegraphics[width=0.7\linewidth]{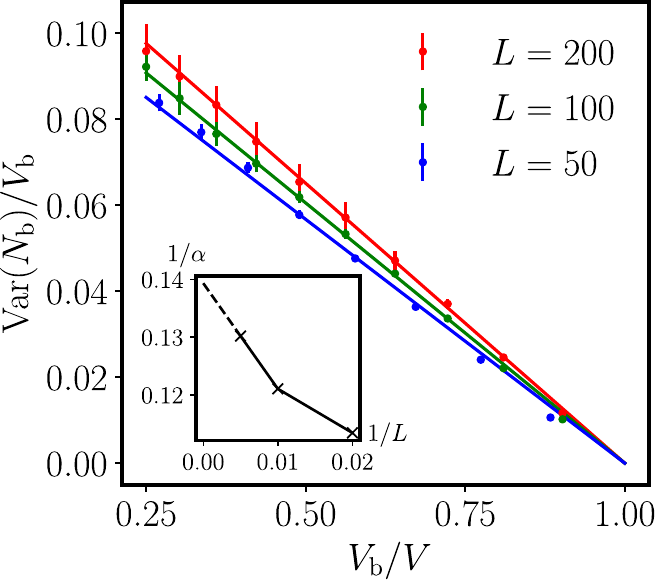}
    \caption{Bulk density fluctuations of the $2d$ Ising model, at $T = 2.0$ and $\phi = 0.04$. Inset: plot of $1 / \alpha$ against $1 / L$. The error bars of the data points in the insets are omitted as they are smaller than the symbol size. The dashed line shows a linear extrapolation.}
    \label{fig:pas_alpha}
\end{figure}

We show a measurement of $\alpha$ in Fig.~\ref{fig:pas_alpha}, where we prepare the system at the mean vapor density extracted from Fig.~\ref{fig:pas_den} (the liquid case will yield the exact same value due to the symmetry of the Ising model). Our procedure produces an estimate $\alpha = 7.27 \pm 0.06$, where the error bar is produced by the comparison between the worst possible extrapolations from available data.

Using the densities and $\alpha$ measured above, we can in principle test the validity of \eqref{eqn:lapcp} and \eqref{eqn:lap}. For the Ising model, \eqref{eqn:lapcp} is trivially satisfied due to its intrinsic symmetry between up and down spins. Thus, we can substitute the $\gamma$ measured through CWT into \eqref{eqn:lap}. With $R$ obtained from the lever rule, we estimate $\rho_l^d - \rho_l^s = \rho_v^d - \rho_v^s = (0.58 \pm 0.01) \times 10^{-3}$. The measured values of the density differences are $\rho_l^d - \rho_l^s = (0.62 \pm 2.03) \times 10^{-3}$ and $\rho_v^d - \rho_v^s = (0.88 \pm 1.12) \times 10^{-3}$. The (large) error bars of the densities are the standard errors of the average density measured on independent CPU threads. Due to the extreme sensitivity of the density difference measurements due to their small values, and the slow convergence in this model, we cannot conclusively conclude on the status of the theory derived in Sec.~\ref{sec:prob}, but the average values of the density differences fall in the right ballpark.  For the ALG, convergence is much faster and density differences are estimated more accurately (smaller relative error).

\section{Density histograms of the phase separated states}
\label{sec:hist}

\begin{figure}
    \centering
    \includegraphics[width=1.0\linewidth]{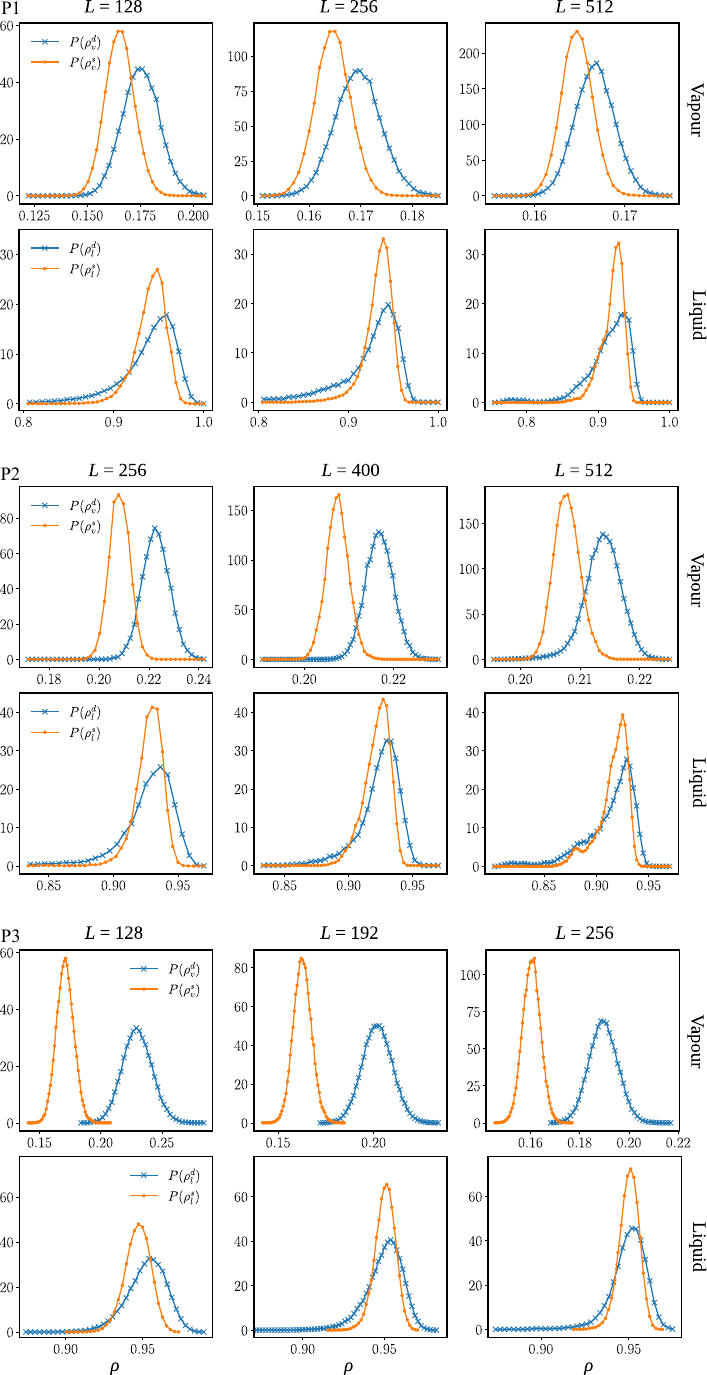}
    \caption{Histograms of densities measured for both the slab and droplet geometries within the vapor and liquid phases. Note that the widths of the histograms are influenced by the box sizes $V_b$ used in these measurements, which are listed in Table~\ref{table:vb}.}
    \label{fig:rho}
\end{figure}

\begin{table}
\centering
\begin{tabular}{| c | c | c | c | c |} 
    \hline
    P1 & $\rho_v^s$ & $\rho_v^d$ & $\rho_l^s$ & $\rho_l^s$ \\
    \hline
    $L = 128$ & 6912 & 4063 & 1536 & 576 \\ 
    \hline
    $L = 256$ & 27136 & 14911 & 6144 & 2500 \\ 
    \hline
    $L = 512$ & 108544 & 59644 & 25600 & 10000 \\ 
    \hline \hline
    P2 & $\rho_v^s$ & $\rho_v^d$ & $\rho_l^s$ & $\rho_l^s$ \\ 
    \hline
    $L = 256$ & 27136 & 14911 & 6144 & 2500 \\ 
    \hline
    $L = 400$ & 68800 & 43036 & 15200 & 5776 \\ 
    \hline
    $L = 512$ & 108544 & 59644 & 25600 & 10000 \\ 
    \hline \hline
    P3 & $\rho_v^s$ & $\rho_v^d$ & $\rho_l^s$ & $\rho_l^s$ \\ 
    \hline
    $L = 128$ & 6144 & 1984 & 1536 & 676 \\ 
    \hline
    $L = 192$ & 13824 & 4464 & 3840 & 1600 \\ 
    \hline
    $L = 256$ & 24576 & 7936 & 6656 & 2704 \\ 
    \hline
\end{tabular}
\caption{The values of $V_b$ used to measure the densities (listed in the column headings) in Fig.~\ref{fig:rho}, for all parameters and system sizes used.}
\label{table:vb}
\end{table}

In Fig.~\ref{fig:rho}, we show the histograms used to extract the data shown in Fig.~\ref{fig:rfit}. We note that the widths of these histograms cannot be directly used compare density fluctuations in different settings, as they all depend on the box sizes $V_b$ used in their extraction. These box sizes are chosen to fit well within the bulk of the phase-separated phases, and are different in each case: see Table~\ref{table:vb}.

\bibliography{bibliography}

\end{document}